%% file: main.tex
\PassOptionsToPackage{dvipsnames,svgnames}{xcolor}

\documentclass[preprint,journal]{vgtc}            

\usepackage{tcolorbox}
\usepackage{multirow}

\vgtcpapertype{system}
\newcommand\bblue[1]{{\color{black}{#1}}}

\title{VizGenie: Toward Self-Refining, Domain-Aware Workflows for Next-Generation Scientific Visualization}

\author{%
  \authororcid{Ayan Biswas}{0000-0002-5535-4549},
  \authororcid{Terece L. Turton}{0000-0003-4345-7783},
  \authororcid{Nishath Rajiv Ranasinghe}{0000-0003-3115-1226},
  \authororcid{Shawn Jones}{0000-0002-4372-870X},
  \authororcid{Bradley Love}{0000-0002-7883-7076}, \\
  \authororcid{William Jones}{0000-0002-3796-6330},
  \authororcid{Aric Hagberg}{0000-0003-2327-3725},
  \authororcid{Han-Wei Shen}{0000-0002-1211-2320},
  \authororcid{Nathan DeBardeleben}{0000-0002-5593-9205},
  \authororcid{Earl Lawrence}{0000-0002-6473-1887}
}

\authorfooter{
  \item
  	A. Biswas, T.L. Turton, N.R. Ransinghe, S. Jones, B. Love, A. Hagberg, N. Debardeleben, E. Lawrence are with Los Alamos National Laboratory.
  	E-mail: {ayan, tlturton, ranasinghe, smjones, love, hagberg, ndebard, earl}@lanl.gov.
  \item
  	William Jones is with Coastal Carolina University.
    E-mail: wjones@coastal.edu.

  \item Han-Wei Shen is with Ohio State University.
  	E-mail: shen.94@osu.edu.
}

\abstract{%
We present VizGenie, a self-improving, agentic framework that advances scientific visualization through large language model (LLM) by orchestrating of a collection of domain-specific and dynamically generated modules. Users initially access core functionalities—such as threshold-based filtering, slice extraction, and statistical analysis—through pre-existing tools. For tasks beyond this baseline, VizGenie autonomously employs LLMs to generate new visualization scripts (e.g., VTK Python code), expanding its capabilities on-demand. Each generated script undergoes automated backend validation and is seamlessly integrated upon successful testing, continuously enhancing the system’s adaptability and robustness. A distinctive feature of VizGenie is its intuitive natural language interface, allowing users to issue high-level feature-based queries (e.g., “visualize the skull” or “highlight tissue boundaries”). The system leverages image-based analysis and visual question answering (VQA) via fine-tuned vision models to interpret these queries precisely, bridging domain expertise and technical implementation. Additionally, users can interactively query generated visualizations through VQA, facilitating deeper exploration. Reliability and reproducibility are further strengthened by Retrieval-Augmented Generation (RAG), providing context-driven responses while maintaining comprehensive provenance records. Evaluations on complex volumetric datasets demonstrate significant reductions in cognitive overhead for iterative visualization tasks. By integrating curated domain-specific tools with LLM-driven flexibility, VizGenie not only accelerates insight generation but also establishes a sustainable, continuously evolving visualization practice. The resulting platform dynamically learns from user interactions, consistently enhancing support for feature-centric exploration and reproducible research in scientific visualization.

}

\keywords{Scientific data, Large language models, Agentic workflows, Natural language, Feature-based visualization.}

\graphicspath{{figs/}{figures/}{pictures/}{images/}{./}} 

\usepackage{tabu}                      
\usepackage{booktabs}                  
\usepackage{lipsum}                    
\usepackage{mwe}                       

\usepackage{mathptmx}                  

\begin{document}


\firstsection{Introduction}

\maketitle

\input{01-introduction}

\section{Related Work} \label{sec:related} 

\input{02-related_work}

\section{System Requirements} \label{sec:sys-reqs} 

\input{03-system_req}

\section{System Overview} 

\input{04-sys_overview}

\section{Methodology and System Design} 

\input{05-methodology}

\section{Interface Design} \label{sec:interface} 

\input{06-interface}

\section{Case Study} 

\input{07-case_study} \label{sec:case-study}

\section{Discussions} 

\input{08-discussion}

\section{Limitations and Future Work} 

\input{09-limitations_future_work}

\clearpage

\acknowledgments{
Research presented in this article was supported by the Laboratory Directed Research and Development program of Los Alamos National Laboratory (LANL) under project number 20250637DI. LANL is operated by Triad National Security, LLC, for the National Nuclear Security Administration of U.S. Department of Energy (Contract No. 89233218CNA000001). This research used resources provided by the Darwin testbed at LANL which is funded by the Computational Systems and Software Environments subprogram of LANL’s Advanced Simulation and Computing program (NNSA/DOE). It is released under LA-UR-25-26045. 

}

\bibliographystyle{abbrv-doi-hyperref}

\bibliography{main}

\end{document}

%% file: 01-introduction.tex
Scientific visualization is essential for deciphering complex, high-dimensional datasets encountered across various scientific disciplines. However, traditional visualization workflows frequently impose considerable cognitive overhead, requiring extensive manual intervention, meticulous parameter tuning, and deep domain-specific expertise. These challenges intensify when dealing with large-scale or volumetric datasets typical in high-performance computing (HPC) contexts, where iterative refinement can become exceedingly burdensome. With ongoing growth in data complexity and volume, there is an urgent need for intelligent, adaptive visualization systems that effortlessly translate high-level, feature-oriented queries into meaningful visual insights, progressively evolving to address new challenges as they emerge.

Recent developments in natural language interfaces and large language models (LLMs) offer promising avenues for automating data visualization processes. Yet existing solutions primarily focus on one-shot code generation approaches, often needing manual corrections or limiting users to fixed, predefined functionalities. While some systems incorporate dynamic code generation and multi-agent orchestration, these approaches frequently struggle to integrate newly generated modules seamlessly, systematically validate visual outputs, or adapt flexibly to 
domain-specific requirements. Current LLM-driven visualization tools predominantly handle simpler 2D charting and dashboard generation, leaving substantial gaps in addressing the complexities inherent in HPC-scale 3D volume rendering and iterative, domain-specific exploration. Unlike existing approaches, VizGenie addresses these critical limitations by integrating robust visual question-answering (VQA) capabilities, rigorous backend validation, and continuous module refinement, significantly enhancing reliability and user trust.

A growing body of AI-driven visualization assistants—including language-to-visualization tools (e.g., Chat2VIS~\cite{maddigan:2023:chat2vis}, ChatVis~\cite{mallick:2025:chatvis}, ncNet~\cite{luo2021_ncNet}), mixed-initiative analytics agents (e.g., LightVA~\cite{zhao:2024:lightva}, PhenoFlow~\cite{kim:2024:PhenoFlow}), and multi-agent chart refinement pipelines (e.g., MatPlotAgent~\cite{yang:2024:matplotagent}, VisPath~\cite{seo:2025:vispath}, PlotGen~\cite{plotgen})—highlights the significant potential of LLM-based automation for visualization tasks. However, the next critical frontier lies in creating self-improving, domain-aware workflows capable of systematically generating, validating, refining, and integrating visualization modules in a continuous, iterative manner. Achieving this requires not only robust code generation capabilities but also mechanisms to ensure reproducibility, manage interaction provenance, and efficiently adapt to sophisticated, evolving visualization needs.
Addressing these shortcomings is not merely incremental but essential, as the complexity and scale of modern HPC datasets demand visualization tools that minimize user intervention while maximizing insight generation and reproducibility.

VizGenie, a novel self-improving agentic framework designed explicitly for adaptive, feature-centric scientific visualization in HPC contexts. Unlike traditional single-pass visualization systems, VizGenie orchestrates a seamless integration of stable, preexisting visualization tools and dynamically generated modules, supported by powerful LLM orchestration. VizGenie continuously enhances its capabilities through a structured process of module generation, rigorous validation, and systematic integration into its visualization repertoire. Users begin their data exploration with foundational tasks—such as filtering, slicing, and basic statistical analyses—but can seamlessly transition to complex, specialized queries that automatically trigger new visualization module generation.

By leveraging advanced natural language querying and domain-specific image-based analysis powered by locally fine-tuned vision models, VizGenie substantially reduces manual overhead. Queries such as “visualize the skull” or “highlight tissue boundaries” are intuitively transformed into actionable visualization scripts, eliminating the need for detailed technical expertise. 
Users can also interactively query and analyze the generated visualizations via natural language, leveraging VizGenie's VQA capabilities to obtain precise, context-specific insights.
This process is further enhanced through Retrieval-Augmented Generation (RAG), providing rich, context-driven insights and ensuring high-quality results. Moreover, VizGenie's built-in memory systematically logs prior interactions, queries, and validated visualizations, facilitating reproducibility and incremental improvement of the visualization pipeline.
By autonomously refining its visual knowledge base and systematically improving VQA precision, VizGenie establishes a sustainable and adaptive visualization workflow that continuously evolves in response to user interactions and data-driven feedback.

Comprehensive evaluations using complex volumetric datasets demonstrate VizGenie’s efficacy and adaptability, significantly accelerating insight generation while establishing new standards for interactive and iterative scientific visualization workflows. The main contributions of this paper include:

\begin{enumerate}[topsep=1pt,itemsep=0pt,partopsep=0pt, parsep=0pt]
    \item \textbf{Agentic Framework Integration:} VizGenie uniquely integrates robust, domain-specific visualization tools with dynamically generated visualization modules, effectively orchestrated by LLMs.

    \item \textbf{Self-Improving Visualization Pipeline:} We introduce a continuous cycle of dynamic visualization module generation, validation, and integration, allowing the system to progressively adapt and expand its capabilities over time.

    \item \textbf{Feature-Based Natural Language Interface:} By combining intuitive natural language interaction with image-based feature analysis, visual question answering, and retrieval-augmented reasoning, VizGenie effectively bridges domain expertise and complex HPC visualization tasks, minimizing cognitive overhead.

    \item \textbf{Rigorous Evaluations and Provenance:} We provide extensive experimental results demonstrating VizGenie’s effectiveness on real-world scientific datasets, underscoring substantial improvements in efficiency, reproducibility, and domain-specific visualization accuracy.
\end{enumerate}

%% file: 02-related_work.tex
The goal of automating data visualization has a long history e.g.~\cite{mackinlay:1986:auto,heer:2009:protovis}. Recently, the advent of AI/ML and natural language interfaces (NLI) has resulted in new ways to automate the visualization process. The success of applying NLI and LLM approaches for automated visualization in information visualization strongly motivates using those capabilities for large-scale scientific data visualization, including simulation data produced by supercomputers.


Scientific data, particularly large-scale simulation data or other 3D data, presents special challenges for automating the visualization process. The features and regions of interest are often amorphous objects that shift and change as the view angle changes. The language used to describe the scientifically interesting features is domain-specific jargon that may be quite different from the natural language on which large language models are trained. Scientific data is often multi-dimensional, temporal, highly evolving, and needs specialized visualization applications~\cite{childs:2012:VisIt, ahrens:2005:paraview, larsen:2022:ISVFCS} and data structures~\cite{vtkBook:2006, harrison:2022:conduit}. For 3D simulation data, the important features may span multiple variables or data types. The appropriateness of different volume visualization techniques~\cite{beyer:2015:visstar, sarton:2023:visstar, xu:2021:surveyvolvis} varies depending on data, compute resources or visualization goals. Techniques such as streamlining or lighting can be used to highlight specific features or emphasize temporal change. Color mapping choices and techniques can be used to accentuate features.  

There is limited research in automating visualizations for scientific data, particularly large-scale HPC-derived simulation data.  ChatVis~\cite{mallick:2025:chatvis} is used to develop ParaView~\cite{ahrens:2005:paraview} Python scripts in an iterative LLM-based workflow that includes a feedback loop that takes resultant error messages and inputs them back into the LLM until the script can correctly generate the requested visualization. 
VizGenie builds upon recent approaches like ChatVis, which also employs LLM-based workflows for scientific visualization. ChatVis focuses primarily on iterative generation and debugging of ParaView scripts through natural language prompts. In contrast, VizGenie expands significantly on this paradigm by integrating domain-aware, feature-centric querying, VQA via fine-tuned vision models, and continuous self-refinement alongside RAG capabilities. Additionally, VizGenie orchestrates a broader array of dynamically generated modules beyond simple scripting, systematically validating and integrating these modules into an evolving agentic workflow.

\begin{figure}[t]
\centering
\includegraphics[width=\linewidth]{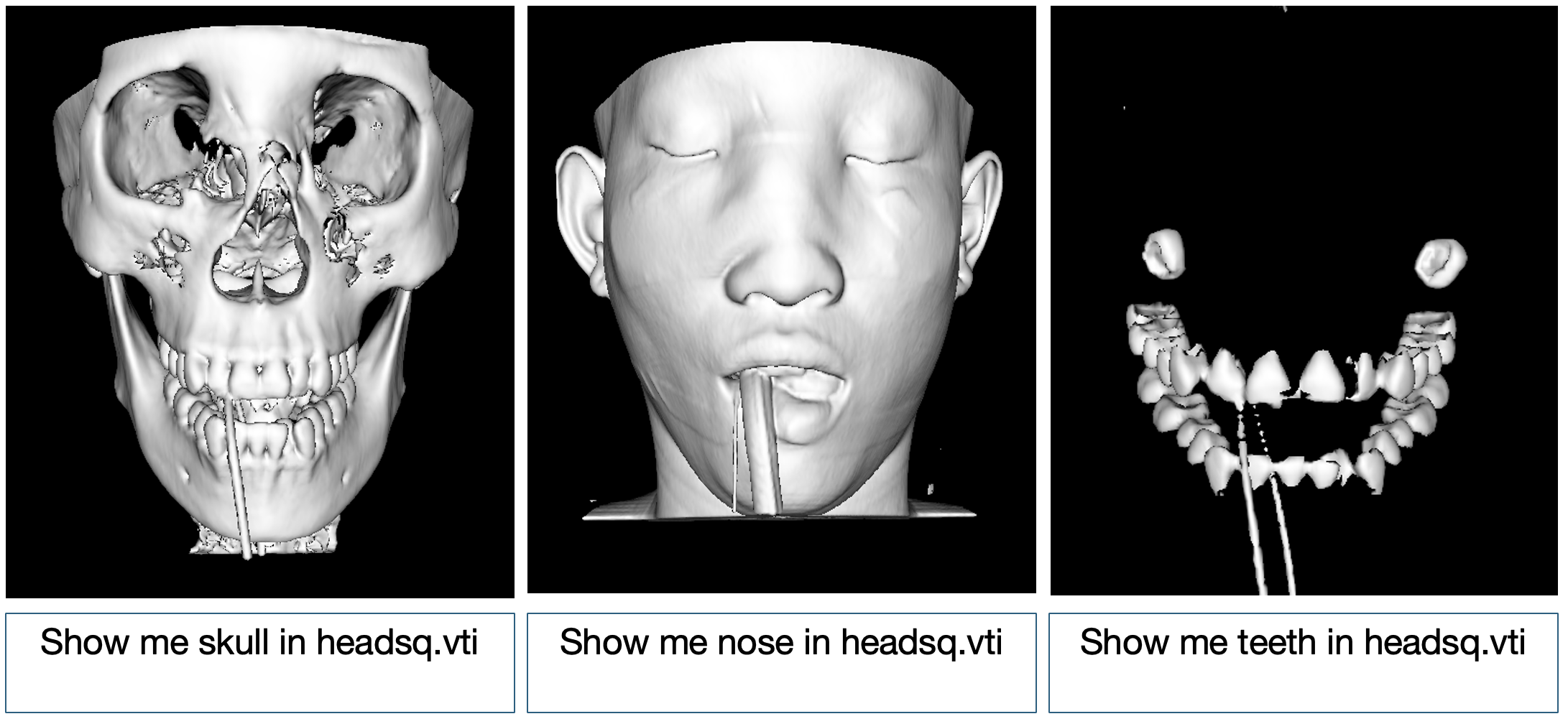}
\caption{An example of feature-based interactions of a user with VizGenie for CT scan data set. Users can make open-ended natural language queries to VizGenie for which predefined tools or functionality are not specified.}
\label{fig:sys_view_teaser}
\vspace{-1em}
\end{figure}

Many LLM-based information visualization workflows have informed this work. Vega~\cite{satyanarayan:2016:vega} and Vega-lite~\cite{satyanarayan:2017:vega-lite} introduced a data flow architecture and a high-level grammar for graphics, encodings, and composing to generate automatic visualizations. NcNet~\cite{luo2021_ncNet} uses a transformer model to go from query to visualization.  By including the capability of one-shot hints or iterative conversation to improve end results, ncNet foreshadows an \bblue{agentic} approach. Chat2VIS~\cite{maddigan:2023:chat2vis} explored the efficacy of several LLMs combined with prompt engineering to produce more effective visualizations.  FREYR~\cite{gallotta:2025freyr} introduces a framework that breaks down tool usage into separate steps to achieve better performance.  
LM4HPC~\cite{chen:2023:lm4hpc} provides a framework for HPC related tasks such as code analysis and optimization. A wide range of approaches can be found in the survey literature~\cite{battle:2021:review, shen:2023:surveynli}.


The field of human-AI collaboration in visual analytics is growing. In this collaborative approach, users interactively guide and modify the visualization process. This research highlights the benefits of integrating natural language interfaces with traditional visualization tools to reduce cognitive overhead and improve overall usability. Recent examples include LightVA~\cite{zhao:2024:lightva} and PhenoFlow~\cite{kim:2024:PhenoFlow}.  LightVA develops a framework with an agent workflow for task planning and execution. PhenoFlow creates a domain-specific framework to support visual analytics in the medical field of stroke management. An LLM is used as the \textit{data wrangler} while the NLI supports the medical expert to interact with visual analytics. 

\begin{figure*}[t!]
\centering
\includegraphics[width=0.9\textwidth]{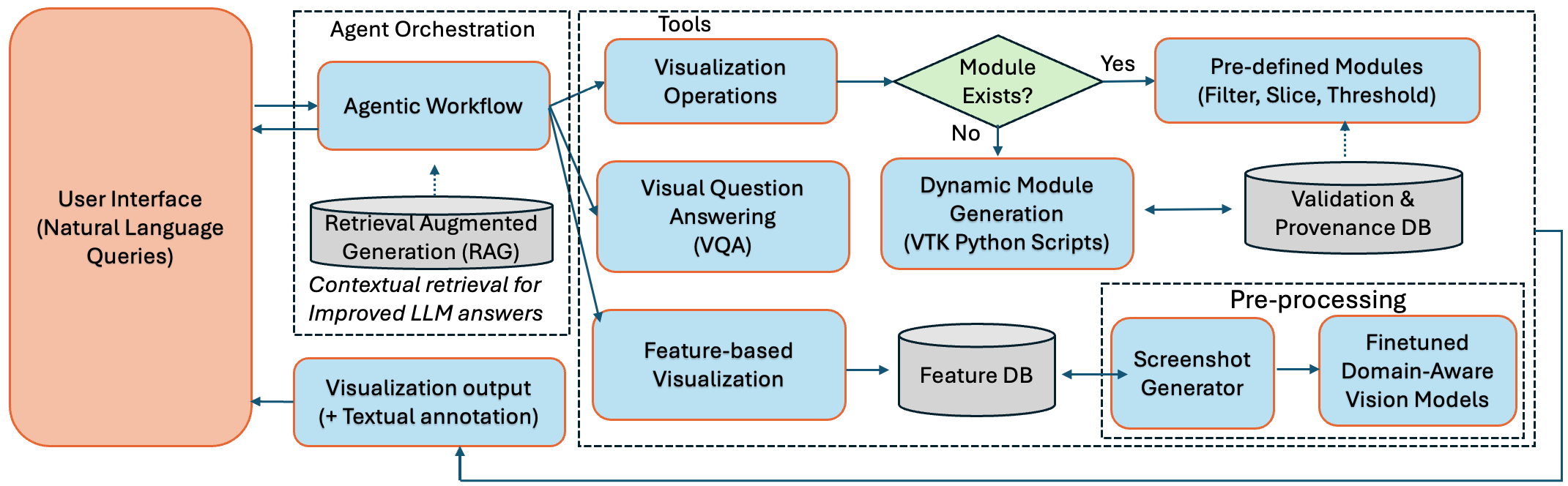}
\caption{\bblue{A schematic diagram of our proposed workflow.}}
\label{fig:workflow}
\vspace{-1em}
\end{figure*}
    
Research such as VisPath~\cite{seo:2025:vispath} and PlotGen~\cite{plotgen} employs multi-agent or ensemble approaches to generate, validate, and refine visualization outputs. These works emphasize coordinated agents that handle different sub tasks (e.g., planning, code generation, and feedback) and demonstrate the benefits of diverse reasoning paths for robust visualization synthesis. Both VisPath and PlotGen leverage MatPlotBench~\cite{yang:2024:matplotagent} for validation.  
Finding adequate benchmarks is a challenge for automated scientific visualization systems. Two commonly used ones for information visualization, MatPlotBench and NVBench~\cite{luo:2021:nvbench}, offer standardized datasets but those datasets and visualizations are geared towards typical information visualization tasks (e.g., bar charts, histograms, etc). Both benchmarks include sets of example visualizations. However, these are primarily 2D plots that do not include the large scale scientific data of the type targeted by VizGenie.  



Several studies incorporate mechanisms for automatic validation, error detection, and iterative improvement of generated visualization modules. These papers address the challenges of maintaining reproducibility and adapting to new data or evolving user requirements, providing a contrast to approaches that rely solely on one-time code generation. 
HaiChart \cite{Haichart}, used Reinforcement learning (RL) framework to iteratively refine and generate high-quality data visualizations using a Monte Carlo Graph Search algorithm \cite{MCGS}, composite reward function, and user feedback, outperforming both human-powered and other AI-powered tools. On the other hand, ShiftScope \cite{shiftscope} uses two actor-critic agents based on the PPO algorithm \cite{ppo} to interactively adapts to users’ evolving data focus by learning to recommend personalized and relevant visualizations.

    
    

Pre-trained vision language models (VLM) such as Llava~\cite{liu:2023:llava}, Llama-3.2-vision, Qwen2-VL~\cite{qwen2vl} models can be efficiently fine tuned to adopt for domain specific sciences and use cases using data centralized implementation applying  Low-Rank Adaptation (LoRA~\cite{hu:2022:lora}) and in federated learning settings using Flora \cite{FLoRA}. LoRA introduces trainable low-rank matrices into the attention projection layers (specifically targeting the query, key, and value matrices), thereby drastically reducing the number of parameters that require updating, reducing computational cost and risk of overfitting while training with small data sets. 
    

%% file: 03-system_req.tex
To effectively address the challenges in modern scientific visualization workflows, particularly for complex and large-scale high-performance computing (HPC) data, we identify the following key requirements:

\begin{description}[leftmargin=!,labelwidth=\widthof{\bfseries R4}] 
    \item[\textbf{R1. Natural Language Interaction:}] The system must interpret high-level, domain-specific queries in natural language to reduce cognitive load and bridge the gap between domain expertise and technical visualization implementation.

    \item[\textbf{R2. Dynamic Extensibility:}] Visualization needs evolve continually; thus, the system must autonomously extend its capabilities through dynamic generation, validation, and integration of new visualization modules.

    \item[\textbf{R3. Domain-Aware Visualization:}] The system should leverage domain-specific knowledge to automatically identify and visualize scientifically relevant features, avoiding generic, non-specific interpretations.

    \item[\textbf{R4. Reproducibility and Provenance:}] It is essential to systematically track and log interactions, visualizations, and decisions made by the system to ensure reproducibility and traceability of scientific findings.

    \item[\textbf{R5. Robustness and Error Handling:}] Generated modules must undergo validation, including automatic detection and iterative correction of errors, to maintain robustness and reliability in diverse scientific scenarios.

    \item[\textbf{R6. Responsive User Interface:}] The system must offer an intuitive and interactive user interface that clearly presents visualization outputs, intermediate reasoning steps, and actionable suggestions to facilitate efficient exploration and iterative refinement.
\end{description}

These requirements set the foundational objectives for our VizGenie system, guiding its design and implementation toward delivering a next-generation visualization experience for scientific research.

%% file: 04-sys_overview.tex
VizGenie is specifically designed to address the requirements, Sec.~\ref{sec:sys-reqs},
providing researchers with an intuitive and adaptive visualization experience. Users interact with VizGenie by simply pointing to a directory containing multiple simulation files. They can then pose natural language questions such as \emph{"How many datasets are available?"}, \emph{"What are the characteristics of this simulation run?"}, or more complex feature-based queries such as \emph{"Where does this specific feature occur across simulations?"} or \emph{"Can you visualize the boundaries of the region of interest?"}. These queries are effortlessly translated by VizGenie into actionable tasks. Figure \ref{fig:sys_view_teaser} illustrates an example interaction, highlighting the system’s response to feature-based user queries.

At its core, VizGenie integrates specialized, domain-specific visualization modules alongside dynamically generated components (R2, R3). Users can initiate data exploration by requesting statistical analyses, extracting informative slices, or filtering datasets based on custom thresholds. As queries evolve, the system leverages LLMs to dynamically generate new visualization scripts—such as volumetric rendering via VTK \cite{vtkBook:2006}, specialized feature highlighting, or iterative refinement of existing visualizations—thereby continuously expanding its capabilities.

The intelligent, agentic workflow of VizGenie is a critical innovation, automating routine visualization tasks while enabling advanced, domain-specific feature extraction and visualization. Complex queries trigger coordinated processes where the system retrieves, analyzes, and visualizes data using tailored, validated techniques (R5). Rigorous validation ensures generated outputs are robust, and successful modules are seamlessly integrated into the existing visualization repertoire, enabling continuous system refinement and evolution.

By combining natural language processing (R1), dynamic module generation, and specialized image-space analysis using vision models, VizGenie significantly reduces cognitive overhead (R6). It empowers users to concentrate on insights and discoveries rather than manual configuration and technical implementation details. The following sections detail how VizGenie's innovative design, iterative validation mechanisms, and responsive user interface collectively establish a new standard for adaptive, feature-centric scientific visualization workflows. A schematic representation of the workflow is presented in Figure \ref{fig:workflow}.

%% file: 05-methodology.tex

The design of VizGenie directly addresses the system requirements outlined in Sec.~\ref{sec:sys-reqs}.
The following subsections detail our methodological innovations and system components that collectively enable our sophisticated, adaptive visualization workflow.

\subsection{Background and Motivation: Automated Agentic Workflows via User Specifications}


\bblue{
In the context of automated AI workflows, an agent\cite{agents_google} can be described as a model-driven orchestrator that combines internal reasoning traces with explicit externally grounded actions. An LLM-based agent can maintain a dynamic context of thoughts, invoke tools or functional APIs, and adapt its strategy based on intermediate results (e.g., using ReAct paradigm~\cite{yao2023reactsynergizingreasoningacting}). Building on a single agent's ability to utilize multiple tools, a multi-agentic system forms a coordinated ensemble of agents-- each with its own state, expertise and sub task focus. These agents can communicate with each other, negotiate or delegate responsibilities, and collaboratively decompose and solve user provided goals. 

Such agentic workflows can offer a far greater flexibility and robustness than static, rule-based automation approaches.}

Inspired by recent advancements in automated agentic systems, we initially developed a purely automated agentic workflow where \bblue{LLM-based agents} dynamically generated and executed code based solely on natural language instructions. \bblue{In this multi-agent workflow: Agent 1, the orchestrator-planner, transforms the user specification into concrete tasks, coordinates their execution, and evaluates intermediate results; Agent 2, the code generator-executor, iteratively produces, runs, and refines code or system commands to meet those tasks; and Agent 3, the security sentinel, scrutinizes every proposed snippet and command for harmful patterns, data exfiltration, or operations outside the permitted environment. To further contain security risks from LLM-based automated workflows, we combined OS isolation with LLM policy filtering. Every agent-generated Python program ran on a network-connected laptop in a single fresh, non-privileged account whose access was confined to a single experiment directory, blocking any change to system files or other users’ data.  

By clearly separating planning, implementation, and protection, this orchestrator–executor–sentinel loop}
provided flexible, just-in-time visualization generation without pre-installed software dependencies. 
However, we quickly observed limitations: relying exclusively on dynamically generated software was computationally inefficient and inconsistent, with roughly half the trials yielding successful outcomes (\bblue{i.e., generated code executed properly and produced visualizations as per users' specifications}). These challenges prompted our development of the hybrid VizGenie system \bblue{where dynamic module generation and execution is coupled with \emph{pre-existing} tools and the system goes through self-improvement cycles independent of user interactions. 
}

\subsection{VizGenie: Visualization-aware Agentic Workflow}

VizGenie integrates stable, predefined visualization modules with the flexibility of dynamic module generation via LLMs. Its natural language interface supports iterative refinement, continuous learning, and 
domain-specific visualization tailored to evolving scientific needs.

\subsubsection{\bblue{Agent Architecture and Tool Selection}}

\bblue{ 
VizGenie's workflow incorporates a single LLM-agent with multiple \textit{tools}. The tools are selected by the LLM agent using \texttt{zero-shot-react-description}. The set of tools can be categorized into two classes: 1) pre-existing (e.g., \texttt{FilterRuns}, \texttt{VisualizeHistogram}, \texttt{VisualizeHistogram}), and 2) dynamic LLM-generated (e.g., \texttt{SimulationInfo}, \texttt{CodeGenerator}, \texttt{ModifyGeneratedCode}, \texttt{LookupFeatureInDataset}).
\textit{Pre-existing} tools enable faster execution and more reliable outcomes. \textit{LLM-generated} modules are dynamically generated when user queries exceed pre-existing list of capabilties (more details in Section \ref{sec:balanced_modules}).

We strategically employ specialized LLMs for distinct phases of our VizGenie workflow—covering both agentic orchestration and the on-the-fly generation of new modules:
\begin{itemize} [topsep=1pt,itemsep=0pt,partopsep=0pt, parsep=0pt]
    \item \textbf{Workflow Orchestration}: A deterministic instance of \texttt{GPT-4o} ensures reliable high-level coordination.
    \item \textbf{Code Generation}: Creative and exploratory code generation is handled by \texttt{o3-mini} with higher temperature settings (temperature=1).
    \item \textbf{Incremental Code Refinement}: For precise modifications, we deploy \texttt{GPT-4o-mini} with deterministic settings to avoid introducing unintended variability and improve response time.
    \item \textbf{General Text Analysis and Q\&A}: General text-based user interactions to handle Q\&A regarding data sets and feature-queries, we deploy \texttt{GPT-4o} for stability and reasonable response time.
\end{itemize}
%
For code generation and modification related tasks, we provide a set of predefined requirements (e.g., \textit{"Generate valid Python code that uses the VTK library to visualize data."}, \textit{"The module must define a function named update\_vtk\_scene(renderer) that takes a VTK renderer as input and updates it."}) and the LLM-agent extracts user-defined specifications (\texttt{prompt}) from the chat interactions and overrides the predefined specs (\textit{"Must Implement these functionalities, overriding the defaults: \texttt{prompt}"}).

}

\subsubsection{Contextualized Data Queries}

\bblue{
VizGenie incorporates context-aware querying through its \texttt{Simulation Info} tool that provides general information about the available data sets. By parsing directory structures of the data folder 
(that has all the data files inside it) and optional user-provided metadata (e.g., README files), the system constructs context-driven prompts based on user queries. These enable the LLM to respond precisely and domain-specifically to question such as "how many data sets do I have?" or "tell me more about xyz.vti data set", significantly enhancing query relevance and reducing ambiguity in exploratory workflows.
}

\subsubsection{Balanced Use of Pre-existing and Dynamic Modules}\label{sec:balanced_modules}

VizGenie combines robust, validated tools (e.g., \texttt{FilterRuns}, \texttt{VisualizeRun}, \texttt{AnalyzeRuns}) with dynamically generated visualization modules created via the \texttt{CodeGenerator} tool. When user queries exceed baseline capabilities, the system autonomously synthesizes new VTK Python scripts using LLMs and \bblue{logs them in a SQLite database that gets validated (details in Section \ref{sec:validate}) for future use. This database acts like a cache that is later looked up by the agent to find modules that match user requirements before it decides to generate new code.} This design ensures both stability and adaptability, meeting evolving visualization demands without compromising reliability.

\subsubsection{Optimizing Code Generation: Full Regeneration versus Incremental Updates}

To optimize computational efficiency and maintain robustness, VizGenie employs a dual strategy for code generation:
\begin{itemize} [topsep=1pt,itemsep=0pt,partopsep=0pt, parsep=0pt]
    \item \textbf{Full Regeneration}: Triggered for entirely new visualization scenarios, ensuring tailored solutions from scratch.
    \item \textbf{Incremental Updates}: Employed for minor adjustments (e.g., color mapping, opacity tuning), minimizing redundancy and preserving validated code integrity.
\end{itemize}
This approach balances flexibility with operational efficiency, significantly reducing error introduction risks.

\subsubsection{Validation and Iterative Error Correction}\label{sec:validate}

To ensure visualization module reliability, VizGenie includes comprehensive validation and iterative correction phases. Newly generated code modules undergo automated execution testing, capturing runtime outputs and errors. Identified issues trigger iterative refinement through targeted LLM prompts until achieving error-free operation or reaching predefined iteration limits. Validated modules are integrated into a persistent cache, ensuring reproducibility and maintaining robust operational standards.
\bblue{
To optimize the response time, our experiment shows that \texttt{o3-mini} produces the best results compared to the other LLMs (such as \texttt{GPT-4o}), reducing the need for iterative evaluation. For ensuring that all the user specifications are included in the code, we perform the full validation via asynchronous self-improvement (Section \ref{sec:self_improve}).
}

\begin{figure}[t]
\centering
\includegraphics[width=0.9\linewidth]{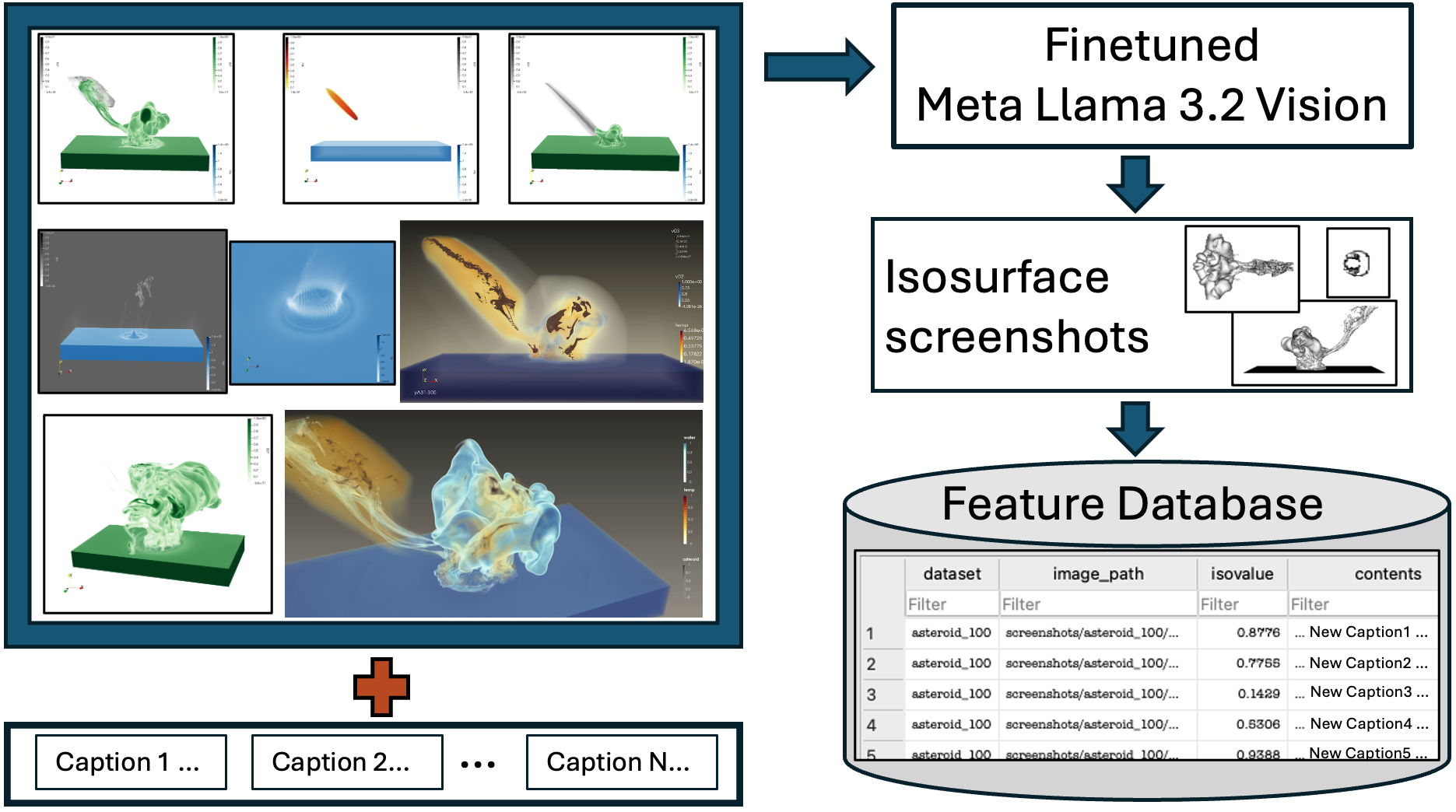}
\caption{A schematic depiction of our fine-tuning workflow for domain-specific adaptation of vision models with Deep Water Impact data set. Images are either drawn from~\cite{patchett:2017:asteroid} or author-generated.}
\label{fig:finetuning}
\vspace{-1em}
\end{figure}

\bblue{\subsubsection{Advanced Feature-based Query Handling Using Domain-Specific Fine-tuning of Llama-Vision Models}}

A core innovation of VizGenie is its sophisticated support for feature-based queries, leveraging a locally fine-tuned Llama-Vision model. The system automates image-based analysis of dataset visualizations, extracting meaningful feature descriptions stored in a structured knowledge base. These metadata significantly improve accuracy in determining optimal visualization parameters (e.g., isovalues) via intelligent, LLM-driven aggregation and reasoning (Figure~\ref{fig:finetuning}). 
\bblue{
Given a data set and a (fine-tuned) vision model, we generate a collection of images for varying camera angles and iso-values that are then used for generating captions via the vision model. These captions, along with the isovalues and data set identifiers are stored in a database. This database is used by the agent to retrieve the most relevant isovalue given the user's query for a given data set, and is visualized. An LLM (\texttt{GPT-4o}) is used to decide the best matching isovalue given the caption descriptions.
}

To accurately support domain-specific queries and enhance the vision model’s understanding of scientific visualization contexts, we implemented a targeted fine-tuning strategy for our locally deployed Llama-Vision model. This fine-tuning process ensures that feature-based queries and VQA outputs are accurate and directly relevant to complex visualization scenarios encountered in scientific exploration.

\paragraph{Dataset Generation:}
We began by systematically generating a representative dataset consisting of screenshots from multiple camera angles and varying visualization parameters (e.g., isovalues, volume renderings with different opacity settings and background colors, etc) for each dataset under consideration.  
In order to test fine-tuning with domain-specific knowledge, we targeted the Deep Water Impact asteroid data
Fig.~\ref{fig:finetuning} with additional images curated from the data release documentation~\cite{patchett:2017:asteroid} and additional visualizations generated from ParaView. The isosurface images ($\approx$ 50 in total) and curated images were accompanied by carefully constructed annotations describing the primary visual features, characteristics, and scientifically relevant observations. One of the authors, a visualization scientist with domain relevant experience, curated, generated, reviewed and refined these image-caption pairs, ensuring their accuracy and relevance.

\paragraph{\bblue{Annotation, Quality Control and Fine-Tuning:}}
Each caption was carefully designed to succinctly describe key visual features, including specific domain terminology and context-specific observations. A  quality assurance process involving  manual reviews by the relevant expert ensured high-quality annotations, minimizing ambiguity and maximizing the dataset’s value for fine-tuning.
\bblue{Next, we utilized parameter-efficient Low-Rank Adaptation (LoRA) \cite{hu:2022:lora} to fine-tune the Llama-Vision model (v3.2 with 11 billion parameters) on a NVIDIA A100-PCIE-40GB GPU.
More details regarding the fine-tuning procedure is available in supplemental material.}





\paragraph{Integration and Evaluation:}
Post-fine-tuning, the enhanced Llama-Vision model was seamlessly integrated back into VizGenie’s workflow. To validate effectiveness, we assessed the accuracy of VQA results on held-out test images and qualitatively examined improvements in the precision of feature descriptions and recommended visualization parameters (e.g., optimal isovalues). Results demonstrated substantial gains in both descriptive accuracy and domain-specific relevance, directly benefiting users in feature-driven explorations. More details about quality analysis are shown in 
Section~\ref{sec:case-study}.  

This targeted domain-specific fine-tuning approach significantly improves the overall capability and reliability of VizGenie’s feature-based query system, enabling 
context-aware visualization for complex scientific datasets.

\subsubsection{Continuous, Asynchronous Self-Improvement}\label{sec:self_improve}

VizGenie employs autonomous, asynchronous refinement mechanisms independent of direct user interaction:
\begin{itemize} [topsep=1pt,itemsep=0pt,partopsep=0pt, parsep=0pt]
    \item \textbf{Improve Toolset:} Periodic validation cycles proactively ensure newly generated modules are error-free and ready for future interactions.
    \item \textbf{Improve VQA Capabilities:} Iterative screenshot analyses refine the underlying knowledge base by identifying and incorporating previously unseen features and metadata.
\end{itemize}
These self-improvement strategies collectively sustain the system's adaptability, responsiveness, and domain relevance.

\subsubsection{Retrieval-Augmented Generation for Contextual Refinement}

To enhance contextual understanding during question/answer sessions, VizGenie implements a lightweight RAG mechanism. By retrieving relevant historical interactions and utilizing domain specific research via locally stored documents, RAG enriches the LLM's contextual awareness, facilitating more accurate default parameter suggestions and improved code quality, particularly when user guidance is minimal. This method elevates the system's usability and domain adaptability.

Collectively, these methodological innovations establish VizGenie as a robust, adaptive visualization platform that systematically addresses the identified system requirements, setting new benchmarks in agentic scientific visualization workflows.

%% file: 06-interface.tex

\begin{figure}
    \centering
    \includegraphics[width=\linewidth]{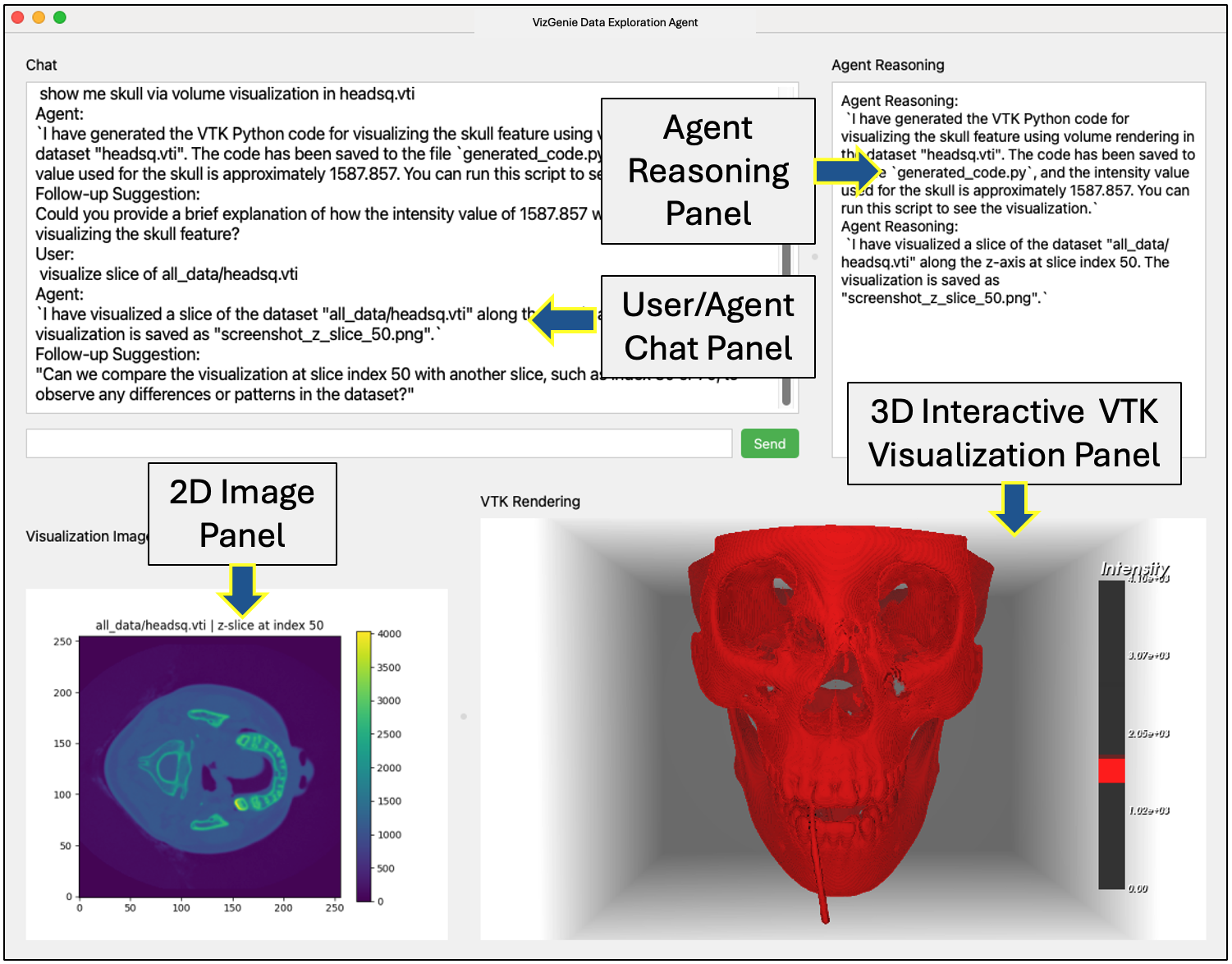}
    \caption{Screenshot of VizGenie's chat and visualization interface.}
    \label{fig:chat_view}
    \vspace{-1em}
\end{figure}

The VizGenie interface is carefully designed to embody and directly support the system requirements in Sec.~\ref{sec:sys-reqs},
providing an intuitive, transparent, and interactive user experience. Developed using PyQt5, the interface integrates various system components seamlessly, enhancing user engagement, reducing cognitive overhead, and ensuring robust domain-specific visualization. Figure \ref{fig:chat_view} illustrates our interactive chat-based user interface, demonstrating how it cohesively supports advanced agentic workflows. \bblue{The interface has four main sections: a user/agent chat panel; an agent reasoning panel; a 2D image display panel; and a 3D VTK visualization rendering panel.  The interface supports interactive callbacks and follow-up suggestions while integrating validation feedback. Details on the interface are in the supplemental material.}

%% file: 07-case_study.tex
All experiments with VizGenie were conducted on a Macbook Pro with M3 chip that has 128GB Memory and 2TB of storage. For our initial case study, we have added four different datasets: 
\begin{enumerate}
  \setlength{\itemsep}{0pt}
  \setlength{\parskip}{0pt}
    \item A CT scan dataset~\cite{ct-head}; this dataset is downloaded with ParaView~\cite{ahrens:2005:paraview} as part of its example data. Originally produced by Siemens Medical Systems, it is a CT study of the head of a twelve year old male with a hole in the skull near the left side of the nose. The volume rendering includes the head and neck from just above the eyes to just below the chin.  
    \item Hurricane Isabel dataset from the 2004 SciVis contest~\cite{hurricane-isabel}; 
    This dataset contains thirteen variables with a resolution of $500 \times 500 \times 100$ for each time step. For our experiments, we have chosen to work with time step 25 and with the pressure field and use subsampled resolution of $250\times250\times50$. 
    \item Astrophysics Turbulence Simulation Data from 2008 SciVis contest~\cite{ion-front}; this data set explores how instabilities in an ionization front related to the formation of the universe's first stars. It contains measurements of the relative abundances of eight chemical species, along with temperature, density, and the velocity field. The data is structured at a resolution of $600 \times 248 \times 248$ per time step, and we have selected time step 99 for our analysis, focusing on the temperature variable.
    \item Deep Water Asteroid Impact dataset from 2018 SciVis contest~\cite{patchett:2017:asteroid}; The Deep Water Impact Ensemble dataset is derived from a series of simulations conducted at Los Alamos National Laboratory to investigate Asteroid Generated Tsunamis. For evaluating our sampling scheme, we selected one ensemble member with a spatial resolution of $300 \times 300 \times 300$. Our analysis focuses on the water volume fraction variable, v02, which ranges from 0.0 (no water) to 1.0 (all water) in the simulation cell. By examining v02, we can visualize the resultant plume of water, thereby facilitating the study of ablation and ejecta as the asteroid impacts the water, producing a plume that spreads into the surrounding area and ascends into the atmosphere. Additionally, a small number of images were curated from the dataset documentation and generated from ParaView with additional variables.  
\end{enumerate}

\begin{figure}[htb!]
    \centering
\begin{tcolorbox}[colback=gray!10, colframe=black, 
sharp corners, boxsep=0pt, left=3pt, right=3pt, top=3pt, bottom=3pt ] 
\small{
\textbf{User:} 
How many datasets do I have? list them with full path
\\ [1ex]
\textbf{Agent:} 
You have four datasets with the following full paths: \\
1. headsq: all\_data/headsq.vti \\
2. isabel\_p\_25\_sub: all\_data/isabel\_p\_25\_sub.vti \\
3. ionization\_front\_0099: all\_data/ionization\_front\_0099.vti \\
4. asteroid\_100: all\_data/asteroid\_100.vti \\[1ex]
Follow-up Suggestion: 
Could you provide a brief description or summary of each dataset to understand their contents and purpose better?
} 
\end{tcolorbox}
\caption{User-Agent Interaction: Data Related Questions }
\label{fig:box-user-agent-data-questions}
\end{figure}

\begin{figure}[htb!]
    \centering
    \begin{subfigure}[t]{0.38\linewidth}
        \centering
        \includegraphics[width=\linewidth]{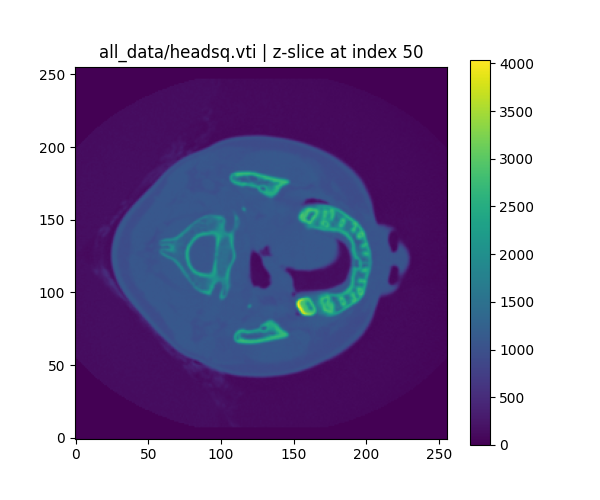}
        \caption{Slice visualization}
        \label{fig:slice}
    \end{subfigure}%
    ~ 
    \begin{subfigure}[t]{0.52\linewidth}
        \centering
        \includegraphics[width=\linewidth]{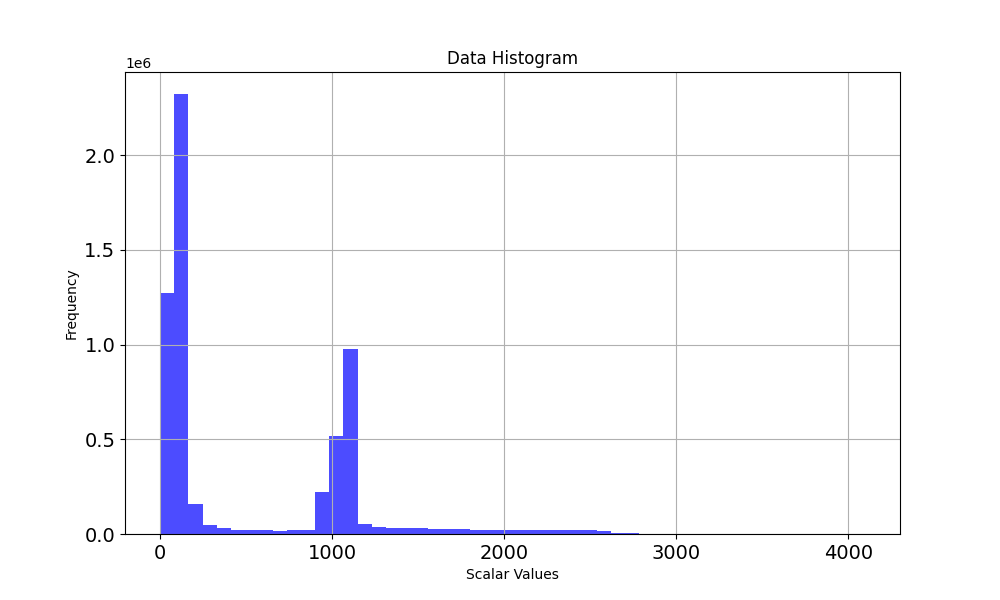}
        \caption{Histogram plot}
        \label{fig:histogram_headsq}
    \end{subfigure}
    \caption{Agent generated images for for CT Scan data (headsq) using preexisting tools as per the user commands as given in Fig.~\ref{fig:box-ctscan}.}
    \label{fig:preexisting_vqa}
    \vspace{-1em}
\end{figure}

We added all the entries in a \textit{.txt} file which serves as the basis for the agent to look up the available datasets. With the initial dataset related interactions \bblue{(Figure \ref{fig:box-user-agent-data-questions})}, now users can go forward with next set of data exploration steps.

\textbf{CT Scan Data Set (headsq.vti)}:
In Fig.~\ref{fig:box-ctscan}, we provide a set of examples for user-agent interactions for the CT scan data set.  The corresponding visualizations are shown in Fig.~\ref{fig:preexisting_vqa}. This also demonstrates the usage of preexisting tools and VQA capabilities of the system using vision models. Llama3.2-Vision model of size 90B was used in the workflow for the VQA examples in Fig.~\ref{fig:box-ctscan}. 

\begin{figure}[htb!]
    \centering
\begin{tcolorbox}[colback=gray!10, colframe=black, 
sharp corners, boxsep=0pt, left=3pt, right=3pt, top=3pt, bottom=3pt ] 
\small{
    \textbf{User:}
    visualize slice of all\_data/headsq.vti\\[1ex]
    \textbf{Agent:}
    The slice visualization of "all\_data/headsq.vti" along the z-axis at index 50 is saved as "screenshot\_z\_slice\_50.png"\\[2ex]
    \textbf{User:}
    what brain structures do you see from screenshot\_z\_slice\_50.png\\[1ex]
    \textbf{Agent:}
    The image "screenshot\_z\_slice\_50.png" likely shows a cross-sectional view of a brain, possibly from an MRI or CT scan. The visible structures include: \\
    1. A central structure resembling the brainstem, identifiable by its characteristic butterfly shape and central canal.\\
    2. Surrounding areas that could be cerebellar hemispheres, based on their location relative to the brainstem.\\
    3. Ventricles or fluid-filled spaces within the brain, possibly the lateral ventricles or parts of the third ventricle, identified by their lower intensity values compared to surrounding tissue.\\[2ex]
    \textbf{User:}
    visualize histogram of all\_data/headsq.vti\\[1ex]
    \textbf{Agent:}
    I have visualized the histogram of the dataset "headsq.vti". The visualization has been saved as "histogram\_plot.png".\\[2ex]
    \textbf{User:}
    for the histogram in histogram\_plot.png, how many modes are there and at what scalar values?\\[1ex]
    \textbf{Agent:}
    There is 1 mode at a scalar value of 0 in the histogram.
} 
\end{tcolorbox}
   \caption{User-Agent Interaction: CT Scan Data. This interaction shows use of preexisting tools and VQA capabilities of the designed system. }
    \label{fig:box-ctscan}
    \vspace{-1em}
\end{figure}

\textbf{Hurricane Isabel Data Set:} For Hurricane Isabel data, we demonstrate a set of user commands and corresponding Agent reasoning in Fig.~\ref{fig:box-isabel}. The resultant visualizations are shown in Fig.~\ref{fig:consecutive_refinement}. Fig.~\ref{fig:box-isabel} shows how the agent can figure out tool-chaining (i.e., calling multiple tools for a given user prompt). Further, we can observe the iterative refinement of the visualizations via successive code-generations and code-modifications.

\begin{figure}[htb!]
    \centering
\begin{tcolorbox}[colback=gray!10, colframe=black, 
sharp corners, boxsep=0pt, left=3pt, right=3pt, top=3pt, bottom=3pt ,
label=box:reasoning]
\small{
    \textbf{User:}
    generate VTK Python code for volume rendering the data set isabel\_p\_25\_sub.vti. Additionally, add a black outline of the data.\\[1ex]
    \textbf{Agent Reasoning:}
    To generate VTK Python code for volume rendering the dataset "isabel\_p\_25\_sub.vti" and add a black outline of the data, I will use the CodeGenerator tool to create the initial code for volume rendering. Then, I will modify the generated code to include a black outline.\\
    First, I will generate the initial VTK Python code for volume rendering.\\
    \textit{Action: CodeGenerator}\\
    \textit{Action Input: \{"prompt": "Generate VTK Python code for volume rendering the dataset.", "data\_path": "isabel\_p\_25\_sub.vti"\}}\\
    \textit{Observation: Code generated and written to generated\_code.py.}\\[1ex]
    Now that the initial VTK Python code for volume rendering has been generated, I will modify the code to add a black outline of the data.\\
    \textit{Action: Modify Generated Code}\\
    \textit{Action Input: \{"modifications": "Add a black outline to the volume rendering.", "code\_file": "generated\_code.py", "output\_file": "generated\_code.py"\}}\\[2ex]
    
    \textbf{User:} change colormap to plasma\\[1ex]
    \textbf{Agent Reasoning:}
    To change the colormap to plasma, I need to modify the previously generated VTK Python code. I will use the Modify Generated Code tool to make this change.\\
    \textit{Action: Modify Generated Code}\\
    \textit{Action Input: \{"modifications": "Change the colormap to plasma.", "code\_file": "generated\_code.py", "output\_file": "generated\_code.py"\}}\\[2ex]
    \textbf{User:}
    modify the code to change the opacity map in such a way: all values up to $-1850$ have high opacity [0.9] and then it linearly goes down to very low opacity [0.001] up to value 0 and for the remaining values, the opacity stays low\\[1ex]
    \textbf{Agent Reasoning:}
    To modify the code to change the opacity map as specified, I need to use the Modify Generated Code tool. This will allow me to adjust the opacity map in the VTK Python code to meet the given requirements.\\
    \textit{Action: Modify Generated Code}\\
    \textit{Action Input: \{"modifications": "Change the opacity map so that all values up to -1850 have high opacity [0.9], then linearly decrease to very low opacity [0.001] up to value 0, and for the remaining values, the opacity stays low."\}}
} 
\end{tcolorbox}
   \caption{User Prompts and Agent Reasoning: Isabel Data }
    \label{fig:box-isabel}
    \vspace{-1em}
\end{figure}

\begin{figure}[thb!]
    \centering
    \includegraphics[width=0.92
    \linewidth]{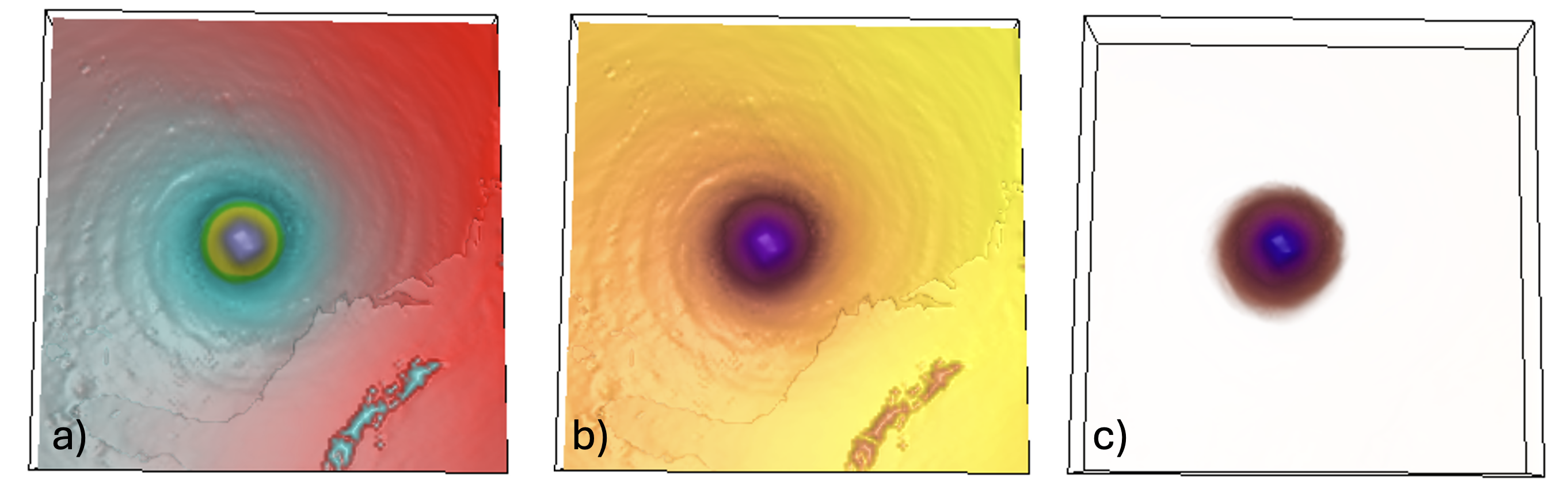}
    \caption{ Results from the sequence of prompts from the user as presented in Fig.~\ref{fig:box-isabel} that enables the user to iteratively refine the visualization parameters to gain scientific insight via code generation and then a set of code modifications [instead of full code regeneration at each step].  }
    \label{fig:consecutive_refinement}
    \vspace{-1em}
\end{figure}


\textbf{Turbulence Data Set:}
For the Turbulence data set,
we show the usage of knowledge base via RAG and how such in-context learning helps add more relevant information to the LLM agents understanding regarding the subject and user's question. For this data set, we added a PDF versions of the SciVis contest webpage~\cite{ion-front} that describes the data and also added a relevant publication by Rosenthal et al.~\cite{rosenthal_2008}. Fig.~\ref{fig:box-ionization} shows an example of a user-agent interaction, showing both answers from without or without RAG support. 
We can observe the LLM answer without RAG is inaccurate. Using RAG, the LLM can generate highly accurate answers.

\begin{figure}[htb!]
    \centering
\begin{tcolorbox}[colback=gray!10, colframe=black,
sharp corners, boxsep=0pt, left=3pt, right=3pt, top=3pt, bottom=3pt,
label=box:ascii_columns]
\small{
\textbf{User:} What are the different fields associated to each of the simulation grid in the IEEE VIS 2008 contest data? Explicitly list the names of all the scalar and vector fields.\\[1ex]
\textbf{Agent (RAG answer):}
The simulation grid contains two types of fields. The scalar fields are::
\begin{enumerate}[topsep=0pt,itemsep=-1ex,partopsep=1ex,parsep=1ex]
    \item \textbf{Temperature} 
    \item \textbf{Mass density} 
    \item \textbf{Chemical species}
\end{enumerate}
In addition, there is one vector field:
    \begin{enumerate}[topsep=0pt,itemsep=-1ex,partopsep=1ex,parsep=1ex]
    \item \textbf{Velocity}\\[1ex] 
\end{enumerate}
\textbf{Agent (Direct LLM answer):}
One acceptable answer was to note that the contest organizers purposely did not “hide” any extra unexpected fields; rather, the simulation was taken from a turbulent‐combustion code and every simulation grid carries exactly the same set of physical fields. In our answer we therefore list the fields that were provided with every grid:

Scalar fields (five in all):
\begin{enumerate}[topsep=0pt,itemsep=-1ex,partopsep=1ex,parsep=1ex]
    \item \textbf{Pressure} 
    \item \textbf{Density} 
    \item \textbf{Temperature} 
    \item \textbf{Mixture fraction (that is, the fuel–concentration field)} 
    \item \textbf{Reaction rate} 
\end{enumerate}

Vector field (one field with three components):
    \begin{enumerate}[topsep=0pt,itemsep=-1ex,partopsep=1ex,parsep=1ex]
    \item \textbf{Velocity} (with its three components, typically labeled u, v, and w)
\end{enumerate}
} 
\end{tcolorbox}
    \caption{User-Agent Interaction for Ionization Front Data shows the usefulness of incorporating RAG as it generally improves accuracy of the LLM-generated answers, in this case, correctly identifying the fields. }
    \label{fig:box-ionization}
    \vspace{-1em}
\end{figure}

\begin{figure}[thb!]
    \centering
    \begin{subfigure}[t]{0.55\linewidth}
        \centering
        \includegraphics[width=\linewidth]{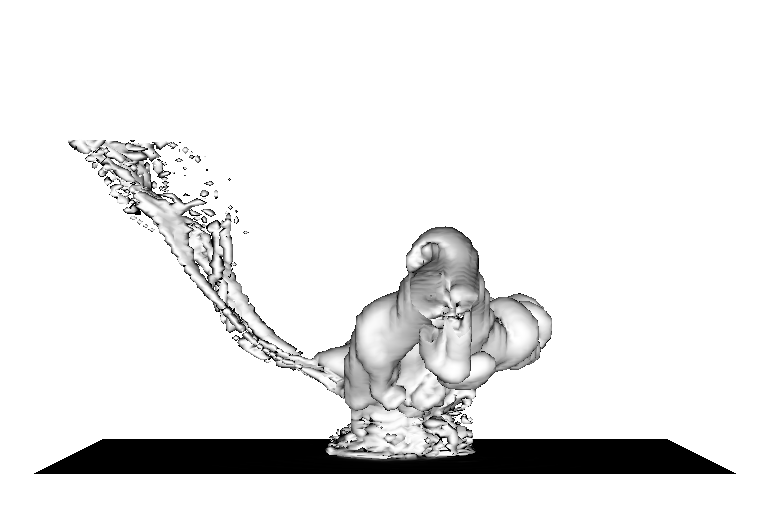}
        \caption{Deep Water Impact data (Isovalue 0.4, VO2 field)}
        \label{fig:asteroid_test}
    \end{subfigure}%
    ~ 
    \begin{subfigure}[t]{0.28\linewidth}
        \centering
        \includegraphics[width=\linewidth]{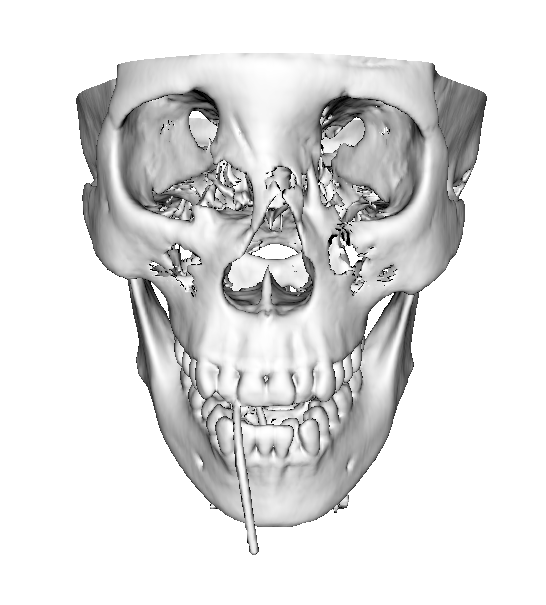}
        \caption{CT Scan data (Isovalue 1500)}
        \label{fig:skull_test}
    \end{subfigure}
    \caption{Isosurface images from (a) Deep Water Impact and (b) CT Scan data (headsq) used for testing fine-tuned vs out-of-the-box vision models as per the user commands as given in Fig.~\ref{fig:box-captions}.}
    \label{fig:finetuning_test}
    \vspace{-1em}
\end{figure}

\textbf{Deep Water Impact Data Set:} For this data set, we demonstrate how domain-specific scientific feature-aware ML models are crucially important for future-looking visualization systems. In Fig.~\ref{fig:box-captions}, we show how our fine-tuned vision model is a big step towards understanding and exploring such domain-specific data sets. When asked to generate a caption for images similar to Fig.~\ref{fig:asteroid_test}, our fine-tuned vision model returns an improved description, compared to the out-of-the-box (OOTB) model that identifies this figure as "cartoon-style" characters. 
Our fine-tuned model still retains general vision capabilities that it had before and we show that by generating captions for Fig.~\ref{fig:skull_test}, where both the models correctly identify this as a human skull.

\begin{figure}[htb!]
    \centering
\begin{tcolorbox}[colback=gray!10, colframe=black, 
sharp corners, boxsep=0pt, left=3pt, right=3pt, top=3pt, bottom=3pt , 
label=box:captions]
\small{
\textbf{User:} Generate caption for Figure \ref{fig:asteroid_test}\\[1ex]
\textbf{Fine-tuned (FT) Model:}
This image captures the moment of impact of a water drop on a free surface. The impact point is at the center of the image, and the direction of the water drop is from the left. The water drop has broken up into a splash, and the water is shown in a transparent gray color...\\[1ex]
\textbf{Out-of-the-Box (OOTB) Model:}
A 3D rendering of a cartoon-style gorilla, its body contorted in a dynamic pose, with its arms and legs bent in a way that suggests intense movement or action. The gorilla appears to be mid-leap, as if it has just launched itself from the ground or is about to land from a jump...\\

\textbf{User:} Generate caption for Figure \ref{fig:skull_test}\\[1ex]
\textbf{Fine-tuned (FT) Model:}
This image shows a 3D rendering of a human skull...\\[1ex]
\textbf{Out-of-the-box (OOTB) Model:}
A 3D image of a skull...
} 
\end{tcolorbox}
    \caption{Generated Captions for Asteroid and CT Scan Data.}
    \label{fig:box-captions}
    \vspace{-1em}
\end{figure}

%% file: 08-discussion.tex
\subsection{VizGenie: Quality Analysis}

\subsubsection{Self-Improvement Through Expanded Visual Database}

To validate the self-improving capability of VizGenie's visual analysis pipeline, we quantitatively assessed the system's knowledge gain resulting from expanding the visual database. Specifically, we compared two scenarios for the \textit{headsq} dataset: a smaller set containing 150 images (captured at 25 isovalues and 6 view angles), and an expanded set with 300 images (captured at 50 isovalues and 6 view angles).

Our analysis showed clear improvements when using the larger image set. Vocabulary richness significantly increased, with vocabulary size expanding from 1606 unique words with 150 images to 2946 unique words with 300 images, reflecting a broader and deeper representation of dataset features. Additionally, semantic analysis revealed that mean pairwise semantic similarity increased from 0.585 (150 images) to 0.717 (300 images). This indicates that with more images, the model achieves greater semantic consistency and a more robust representation of domain-specific features, which directly translates into improved accuracy and relevance of visual question answering.

\begin{figure}[htbp]
    \centering
    \includegraphics[width=0.85\linewidth]{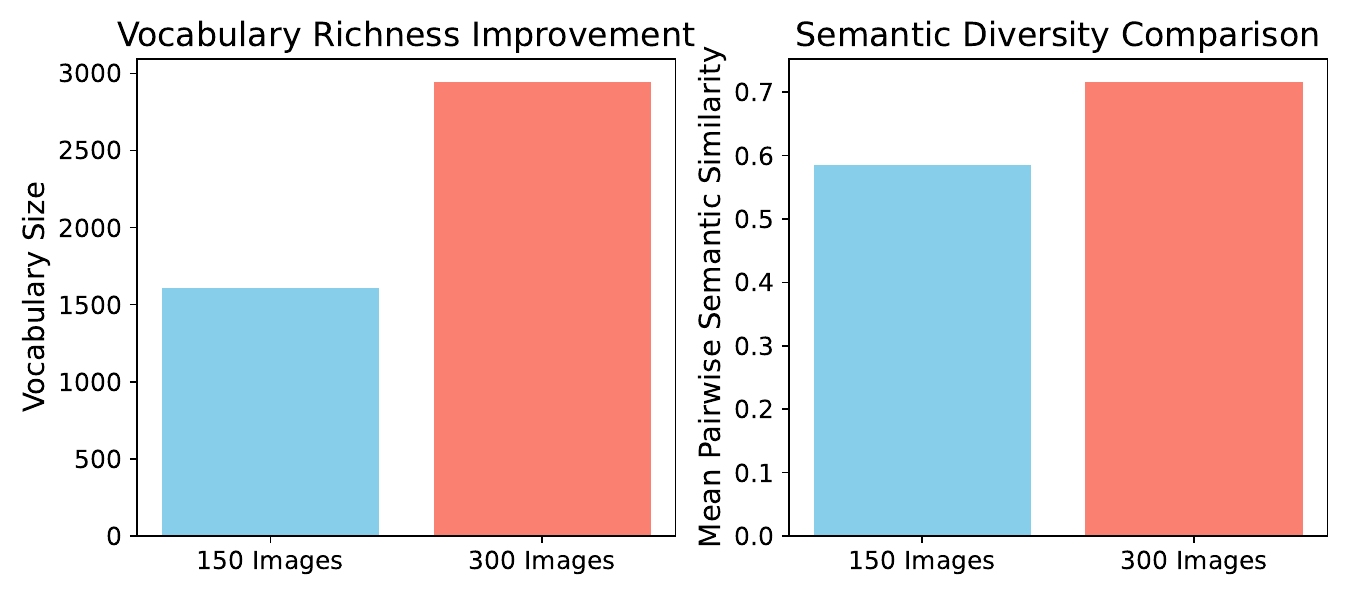}
    \caption{Quantitative analysis illustrating improvements in vocabulary richness and semantic consistency resulting from expanding the visual database from 150 to 300 images for the \textit{headsq} dataset.}
    \label{fig:self_improvement_headsq}
    \vspace{-1em}
\end{figure}

Figure~\ref{fig:self_improvement_headsq} illustrates these improvements, underscoring the effectiveness of VizGenie's asynchronous self-improvement mechanism. By autonomously generating additional images and integrating them into its visual knowledge base, VizGenie systematically enhances its capability for comprehensive, insightful exploration of complex scientific datasets.

\begin{figure}[htbp]
    \centering
    \includegraphics[width=0.87\linewidth]{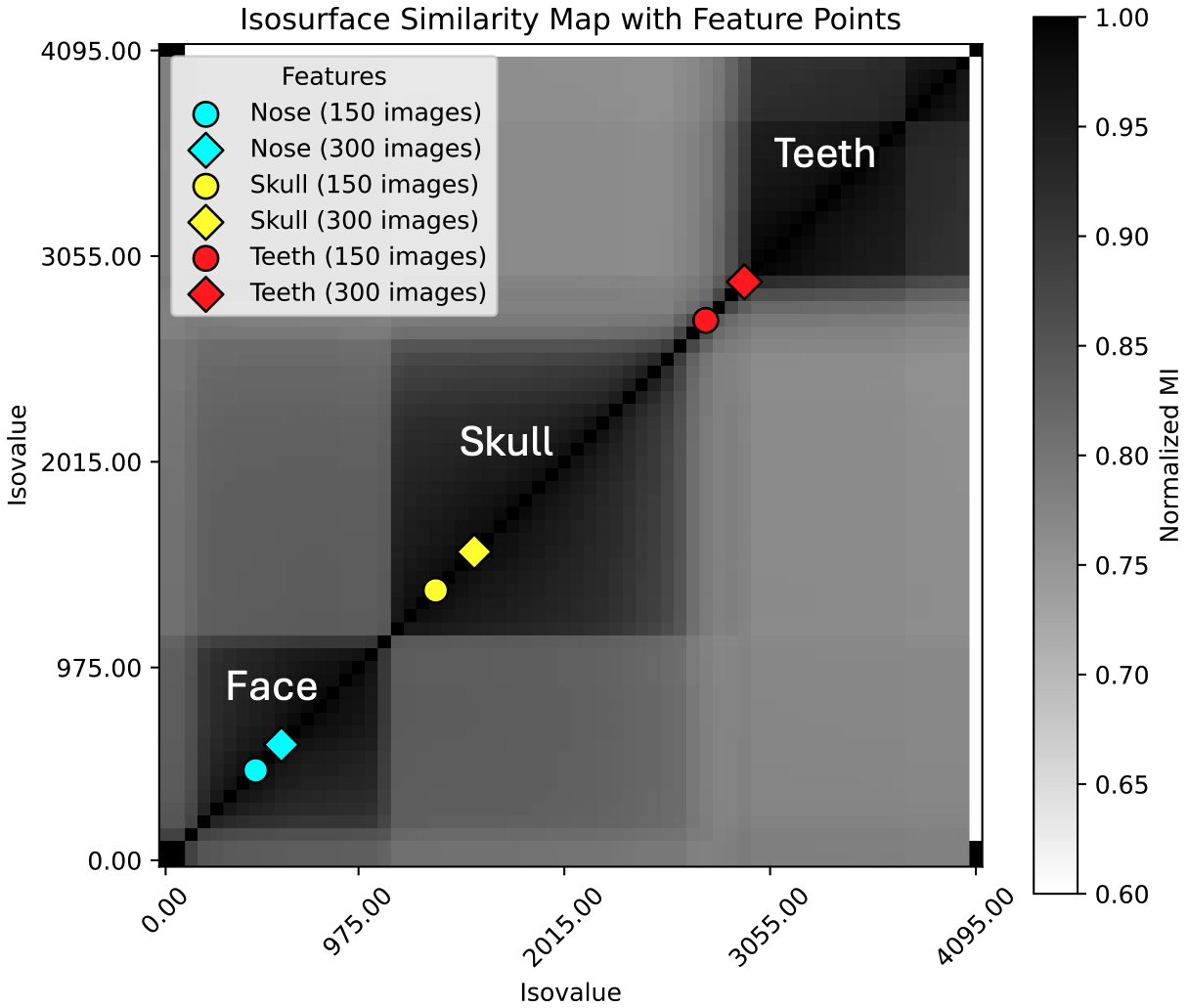}
    \caption{Isosurface similarity map and feature placement changes resulting from expanding the visual database from 150 to 300 images for the \textit{headsq} dataset [high-level feature areas are labeled].}
    \label{fig:iso_sim_headsq}
    \vspace{-1em}
\end{figure}

To further illustrate VizGenie's self-improving capability, we examined changes in predicted optimal isovalues for specific anatomical features—namely, nose, skull, and teeth—before and after expanding the image database from 150 to 300 images. Figure~\ref{fig:iso_sim_headsq} shows the isosurface similarity map (based on normalized mutual information (MI) similarity, by Bruckner and Möller\cite{bruckner_iso}) with marked optimal isovalues as returned by the VizGenie system, for 150 and 300 images. Notably, the predicted optimal isovalues \bblue{aligned more with the ideal isosurfaces as judged by a visualization expert: the nose isovalue moved from 455 to 585 (ideal range is 650-700)}, skull from 1365 to 1587 \bblue{(ideal range is 1550-1600)}, and teeth from 2730 to 2925 \bblue{(ideal range is 2800-2900)}. These shifts underscore the system's enhanced understanding of the dataset, driven by its automatic, iterative expansion of feature understanding.

\subsubsection{Validation of Vision Model Fine-tuning}

To assess the effectiveness of our \bblue{domain-specific fine-tuning} for the Llama-3.2-Vision model (11B parameters), we conducted a comparative analysis of caption generation before and after the fine-tuning procedure. This qualitative analysis was performed on captions \bblue{with newly generated images} from the deep-water asteroid impact dataset. This consisted of 300 screenshots generated across 50 isovalues and 6 camera angles for the $v02$ variable, the volume fraction of water.  These images were supplemented by author-captioned images from the open-source asteroid data set release documentation~\cite{patchett:2017:asteroid} and captioned ParaView visualizations generated by that author.  This supplemental sets had additional variables including the volume fraction of asteroid ($v03$), the density (rho) and the temperature (tev). A held-out set of ten ParaView visualizations were used for this analysis. These new visualizations used the same variables and color maps as the supplemental set of images described previously. 

Our analysis revealed meaningful improvements post fine-tuning. Pre-tuning, the model returned captions that guessed the context: \textit{shockwave emanating from a cylindrical charge underwater}, \textit{water impact by a droplet onto a solid surface}, \textit{dolphin and its jumping behavior in the ocean}, \textit{green, translucent, complex object, possibly a flower}.  The most accurate sentence was: \textit{A 3D simulation of a projectile penetrating a surface, with the projectile's path and the surface's deformation visible}.  
The model before tuning did make some accurate observations about the color map used (green, gray, etc.) and legends. Other than these types of purely observational remarks, there were no accurate captions.

\begin{figure}[htbp]
    \centering
    \includegraphics[width=0.7\linewidth]{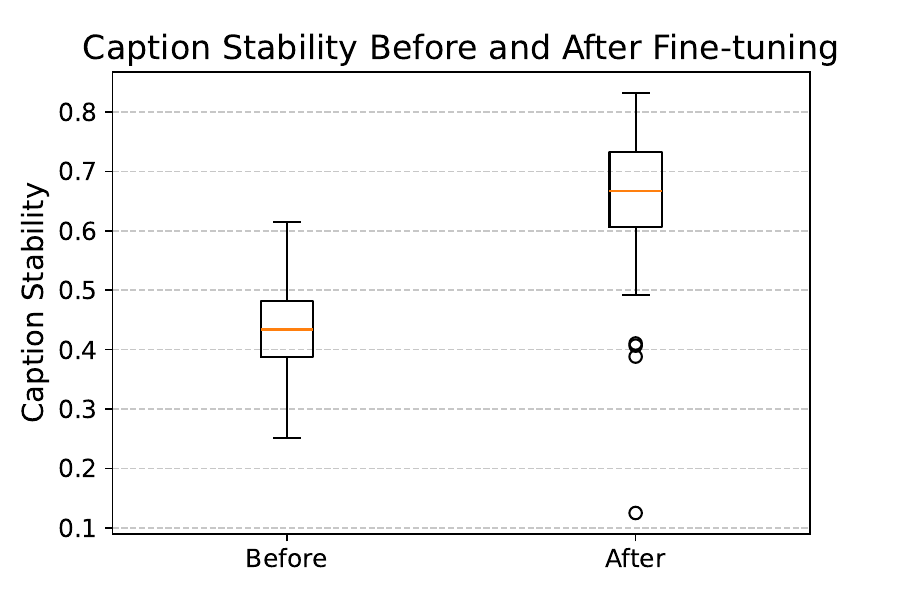}
    \caption{Caption stability (inter-caption similarity) comparison before and after fine-tuning. Stability significantly increased after fine-tuning, reflecting improved consistency and domain specificity in generated captions.}
    \label{fig:caption_stability}
    \vspace{-1em}
\end{figure}

The fine-tuned model produced captions with increased semantic precision and enhanced use of domain-specific vocabulary. Total vocabulary size decreased from 4135 words before fine-tuning to a more concise and focused 2287 words afterward. We observed substantial increases in relevant domain-specific keyword frequencies: occurrences of "crater" rose dramatically from 1 to 348, "asteroid" mentions increased from 0 to 1112, and "impact" rose from 38 to 1616. Conversely, general terms less relevant to the dataset context, such as "shockwave," decreased, demonstrating the model's improved contextual accuracy.  
Additionally, we evaluated caption stability-the consistency of descriptions generated for identical isovalues viewed from different camera angles. After fine-tuning, caption stability increased significantly, with the mean inter-caption semantic similarity, Fig.~\ref{fig:caption_stability}, rising from 0.432 to 0.654 (Mann-Whitney U test, $p < 0.00001$)  

\textbf{Human-in-the-loop Qualitative Analysis} 
For a qualitative analysis of the captions from the fine-tuned model, each separate sentence or phrase in the overall caption was rated by the domain relevant author as TRUE or FALSE. Compared to the effectively 0\% true statements about the images before tuning, the model improved to 39\% accurate statements.  While that percentage seems low, the incorrect statements highlight the difficulties posed for automating scientific visualization.  The jargon for the asteroid impact includes the terms: \textit{water jet}, \textit{water plume}, and \textit{water vapor plume}. While those terms apply to very specific and different phenomena, the model had a tendency to use those terms interchangeably. All color maps use were simple linear color palettes: linear blue (rho), linear green (v02), grey (v03) and yellow-orange-red (tev).  The model often stated incorrect information about the relevant color map such as identifying it as a log scale, when it was not. The relevant term, \textit{transient crater}, was often followed by \textit{basin}, although basin was never used in the training captions and is not a correct description of the phenomenon.  

The multi-variable aspect of some of the image data also contributed to the tendency to produce incorrect statements. The fine-tuned model did not always correctly identify the variable of interest.  Nor could it always identify which physical phenomenon was obscured or, conversely, which variable was doing the obscuring. 

\bblue{\textbf{LLM as a Judge Quantitative Analysis} }
\bblue{As a quantitative study, we used an LLM-as-a-judge approach, prompting the LLM to do a comparison of the out-of-the-box (OOTB) and fine-tuned models  to the ground truth  caption provided by the domain relevant author.  We used GPT-4o and o4-mini-high, a reasoning model, as the judges.  GPT-4o ranked the fine-tuned model captions substantially higher with a mean of 65\% compared to the OOTB model averaging 1.7\%.  The o4-mini-high model was more conservative in its opinion, with an average of 38\%, still higher than the OOTB model at 20\%.  Details are in the supplementary material. 
}

These improvements confirm that the fine-tuned model generates more accurate and consistent captions and enhances the overall 
domain relevance of visual question answering within VizGenie.  However, these results also show that fine-tuning for more natural language visual phenomena such as skull and teeth is more straightforward than fine-tuning for the more amorphous features and technical jargon of a complex large-scale simulation dataset. There are lessons here that need to be further explored such as how many captioned images are needed to fine-tune the model or how to improve model understanding of the scientific jargon, e.g., by adding additional background understanding via domain-specific RAGs.  

\input{performance-table}

\subsection{Timing and Performance}

Table~\ref{tab:llm_performance} summarizes the performance metrics for various visualization tasks executed by different LLM models within VizGenie. Due to the general stochasticity of the LLMs, these numbers are collected by averaging over five executions of the same prompts. 
\bblue{
During this two step evaluation, first we checked if the LLMs are generating executable code and then, if the code executes and produces output, we checked if the user specifications were correctly incorporated. User specifications include visualization choice (e.g., volume rendering, isosurface), visualization parameters (e.g., choice of isovalue, colormaps, background color).
}
We observe that code generation tasks, such as volume rendering and isosurface creation, typically require more processing time (approximately 15 to 23 seconds), whereas simpler operations via preexisting tools like slicing and histogram generation complete significantly faster (about 2 to 4 seconds). Notably, the \texttt{gpt-4o} and \texttt{o3-mini} models generally produce valid and robust outputs (validity scores close to 1), though the smaller (and cheaper) model variants (\texttt{gpt-4o-mini}) occasionally compromise validity but are generally slightly faster. 
For using preexisting tools via agentic reasoning, the \texttt{gpt-4o} model provides the highest quality in reasonable interaction time.
These results highlight a trade-off between model complexity, validity, and response time, underscoring the importance of model selection tailored to task complexity within an interactive visualization workflow.

\subsection{User Feedback}

The authors represent broad expertise from physics, HPC, visualization, and AI/ML. Various  authors contributed labeled input to the framework and user feedback on the interface and case studies.  Most of the authors providing feedback were regular users of LLMs in their daily work.  As early users of the framework, the authors explored various prompts using the chat interface and provided feedback on additional features that would be helpful in the tool. The natural language-based feature exploration and simplicity of the user interface were considered strengths by these early users. The interactive nature of the agent, particularly in suggesting next queries was also well-received. The VQA aspect was highly rated although early users were somewhat disappointed when the VQA system did not yield the exact answer. Early users also noted that at times their prompts were misinterpreted by the agent/LLMs. One primary reason was that vocabulary differs across domains. Keywords such as "show", "explore", "visualize" can mean different operations based on the user and their expertise/understanding of the problem. \bblue{These early use cases informed  the development and evolution of VizGenie.} 

\textbf{Iterative Improvements Driven by User Feedback:}
Early interactions with domain experts and initial users highlighted the need for more accurate and domain-specific visual interpretations, as generic vision models frequently produced irrelevant or misleading descriptions of complex scientific visualizations. 
In response to this critical feedback, we implemented targeted fine-tuning of the vision models (specifically Llama-Vision) using domain-specific datasets. 
This iterative enhancement significantly improved the semantic precision, caption stability, and domain relevance of VizGenie's VQA capability, directly addressing user-identified limitations. Despite the improvements due to fine-tuning, since we use LLM-based operators to search for features in the feature database, we sometimes get unexpected results and must resubmit the prompt to regenerate answers. We would like to address in our future work.

%% file: performance-table.tex
\begin{table*}[t]
\centering
\scriptsize
\caption{Performance summary of LLMs across visualization tasks.}
\label{tab:llm_performance}
\resizebox{0.95\linewidth}{!}{%
\begin{tabular}{|l|c|c|c|c|l|c|c|c|}
\hline
\multirow{2}{*}{\textbf{LLM}} & \multicolumn{2}{c|}{\textbf{Code}} & \multirow{2}{*}{\textbf{Agent}} & \multirow{2}{*}{\textbf{Vis. Task}} & \multirow{2}{*}{\textbf{Prompt (Shortened)}} & \multirow{2}{*}{\textbf{Validity}} & \multicolumn{2}{c|}{\textbf{Time (s)}} \\
\cline{2-3} \cline{8-9}
& \textbf{Gen.} & \textbf{Mod.} & & & & & \textbf{Avg.} & \textbf{Std.} \\
\hline
o3-mini & YES & NO & gpt-4o & Vol. Render & vtk volume rendering (headsq.vti) & 1 & 22.38 & 6.61 \\
gpt-4o-mini & NO & YES & gpt-4o & Vol. Render & change colormaps to plasma & 0.9 & 24.00 & 4.18 \\
o3-mini & NO & YES & gpt-4o & Vol. Render & change colormaps to plasma & 1 & 21.04 & 3.08 \\
o3-mini & YES & NO & gpt-4o & Isosurface & vtk isosurface (isovalue=1500) headsq.vti & 1 & 22.82 & 0.90 \\
o3-mini & NO & YES & gpt-4o & Isosurface & modify isovalue to 500 & 1 & 20.55 & 5.61 \\
gpt-4o-mini & NO & YES & gpt-4o & Isosurface & modify isovalue to 500 & 1 & 19.58 & 3.10 \\
gpt-4o & YES & NO & gpt-4o & Vol. Render & vtk volume rendering (headsq.vti) & 1 & 15.06 & 2.45 \\
gpt-4o-mini & YES & NO & gpt-4o & Isosurface & vtk isosurface (isovalue=1500) headsq.vti & 0 & 18.00 & 2.92 \\
- & NO & NO & gpt-4o & Slicing & slice visualization (headsq.vti) & 1 & 3.20 & 0.45 \\
- & NO & NO & o3-mini & Slicing & slice visualization (headsq.vti) & 0.6 & 16.80 & 2.59 \\
- & NO & NO & gpt-4o-mini & Slicing & slice visualization (headsq.vti) & 0.6 & 3.40 & 0.55 \\
- & NO & NO & gpt-4o & Histogram & histogram (headsq.vti) & 1 & 2.88 & 0.24 \\
- & NO & NO & o3-mini & Histogram & histogram (headsq.vti) & 0.8 & 18.20 & 9.73 \\
- & NO & NO & gpt-4o-mini & Histogram & histogram (headsq.vti) & 0.8 & 2.22 & 0.44 \\
\hline
\end{tabular}%
}
\end{table*}

%% file: 09-limitations_future_work.tex
Despite its promising performance and innovative design, VizGenie has several limitations that provide avenues for future research. The reliance on multiple large language models (LLMs) introduces latency and computational overhead, potentially affecting real-time responsiveness with large datasets or complex queries. Moreover, the accuracy and reproducibility of dynamically generated modules and incremental code modifications depend significantly on LLM outputs, which may occasionally contain errors, variability, or biases originating from training data. \bblue{The current domain-specific fine-tuning approach is manually initiated, limiting seamless generalization across different scientific domains.} Additionally, the retrieval-augmented generation (RAG) mechanism is presently limited, restricting its ability to leverage fully contextual learning for extensive validation or optimal parameter suggestions. 
Further, we currently limit the pipeline to VTK for reading and visualizing the data, directing the LLMs to generate only VTK code. \bblue{Using VTK keeps VizGenie lightweight during this early development process. In the future, we could consider merging VizGenie capabilities with visualization tools such as ParaView or VisIt.}

Future research includes developing more efficient, scalable LLM architectures, such as model distillation and quantization, to reduce latency without compromising output quality. \bblue{We aim to automate domain-specific fine-tuning through continuous learning frameworks, improving VizGenie's adaptability across diverse scientific contexts and data types.} Advanced retrieval-augmented generation methods incorporating sophisticated vector databases and embedding-based retrieval will also be explored to provide richer contextual insights and reduce manual interventions. Beyond VTK functionalities, we will prototype more sophisticated multi-agent systems to auto-analyze and generate new data readers and user-directed non-VTK visualizations.

We will investigate emerging multi-modal LLMs and generative diffusion models to expand visualization capabilities, 
with the goal of
automated novel visualization generation directly from user queries. Extensive user studies \bblue{are needed along with} rigorous quantitative evaluations across diverse real-world datasets. \bblue{These evaluations} will guide VizGenie's iterative refinement, ensuring its sustained relevance and effectiveness within high-performance computing scientific visualization workflows.